\documentclass[A4paper,11pt]{article}
\usepackage{amssymb}
\usepackage[dvips]{graphicx}
\usepackage{color}
\usepackage{amsfonts,amsmath,epsfig}
\usepackage{amsmath, amsthm}
\newlength{\bredde}
\def\slash#1{\settowidth{\bredde}{$#1$}\ifmmode\,\raisebox{.15ex}{/}
\hspace*{-\bredde} #1\else$\,\raisebox{.15ex}{/}\hspace*{-\bredde} #1$\fi}
\theoremstyle{plain}
\newtheorem{thm}{Theorem}[section]

\newtheorem{prop}[thm]{Proposition}
\theoremstyle{definition}

\theoremstyle{remark}

\newcommand{\be}{\begin{equation}}
\newcommand{\ee}{\end{equation}}
\newcommand{\bea}{\begin{eqnarray}}
\newcommand{\eea}{\end{eqnarray}}

\newcommand{\al}{\alpha}

\newcommand{\sect}[1]{\setcounter{equation}{0}\section{#1}}

\begin{document}

\begin{center}
\centerline{\bf  REAL ROOTS OF RANDOM POLYNOMIALS:}
\centerline {\bf UNIVERSALITY CLOSE TO ACCUMULATION POINTS}

\vskip 0.5cm \centerline{ \large \bf Anthony P. Aldous and Yan V.
Fyodorov} \vskip 0.3cm

\centerline{$^2$Department of Mathematical Sciences,
Brunel University}
\centerline{Uxbridge, UB8 3PH, United Kingdom}
\end{center}

\vskip 3cm

\begin{abstract}
We identify the scaling region of a width $O(n^{-1})$ in the
vicinity of the accumulation points $t=\pm 1$ of the real roots of
a random Kac-like polynomial of large degree $n$. We argue that
the density of the real roots in this region tends to a universal
form shared by all polynomials with independent, identically
distributed coefficients $c_i$, as long as the second moment
$\sigma=\mbox{\bf E}(c_i^2)$ is finite. In particular, we reveal a
gradual (in contrast to the previously reported abrupt) and quite
nontrivial suppression of the number of real roots for
coefficients with a nonzero mean value $\mu_n=\mbox{\bf E}(c_i)$
scaled as $\mu_n\sim n^{-1/2}$.
\end{abstract}

\vfill
\thispagestyle{empty}

\newpage

\renewcommand{\thefootnote}{\arabic{footnote}}
\setcounter{footnote}{0}

\sect{Introduction}\label{intro}

Recently, there was an essential resurrection of interest in
statistical properties of zeros of random polynomials and random
analytic functions \cite{bog1}--\cite{Shif2}.

The distribution of real roots for random polynomials was studied
in classical papers by Bloch and Polya~\cite{Bloch}, Littlewood
and Offord~\cite{Lit}, Kac~\cite{Kac} and many others. Many
results and an overview of the field is given in the recent
monograph by Farahmand \cite{Far}.

As is well-known, for the case of algebraic polynomials
\begin{equation}\label{defpol}
f_n(t)=c_0+c_1 t+...+c_{n-1} t^{n-1}
\end{equation}
where the real coefficients $c_i$ are independently and
identically distributed around mean zero, the roots are mostly
located near $\pm 1$. A natural intuitive picture for such an
accumulation was proposed by Kac. Indeed, if all coefficients
$c_i$ are of the same order then only for $|t|=1$ will the terms
have a good chance to interact and cancel each other producing
zero.

In particular, for Gaussian-distributed coefficients
 with mean $E [c_j]=0$
and variance $E[c_j^2]=\sigma$, the mean density of real zeros of
an algebraic polynomial $p_n(t)$ is given by the Kac formula
\begin{equation}\label{Kac}
p_n(t)=\frac{\sqrt{A_n(t)C_n(t)-B_n^2(t)}}{\pi A_n(t)}
\end{equation}
where
\[
A_n(t)=\sigma\sum_{j=0}^{n-1}
t^{2j}=\sigma\frac{t^{2n}-1}{t-1}\,,\quad
B_n(t)=\frac{1}{2}\frac{d}{dt}A_n(t)\,, \quad
C_n(t)=\frac{1}{4}\frac{d^2}{dt^2}A_n(t)+\frac{1}{4t}\frac{d}{dt}A_n(t)\,.
\]

Usually, one is interested in the limit of polynomials of high
degree $n$. For a fixed $t$ the density is given by:
\begin{equation}\label{limtriv}
\lim_{n\to\infty}p^{(f)}_{n}(t\ne \pm 1)=p^{(f)}(t)=
\frac{1}{\pi|1-t^2|}
\end{equation}
and
\[
p_n(\pm 1)=\frac{1}{\pi}\left[\frac{n^2-1}{12}\right]^{1/2}
\]
The expected number of real zeros is known to have the following
asymptotics:
\begin{equation}\label{Kacasy}
N_n(-\infty,\infty)=\int_{-\infty}^{\infty}
p_n(t)dt=\frac{2}{\pi}\log {n}+C+O(n^{-2})
\end{equation}
where the leading logarithmic term was found by Kac himself, and
the corrections, in particular, the constant
\begin{equation}
C=\frac{2}{\pi}\left\{\int_0^{1}\sqrt{\frac{1}{v^2}-
\frac{1}{\sinh^2{v}}}\,dv-\int_1^{\infty}\left[\frac{1}{t}-\sqrt{\frac{1}{v^2}-
\frac{1}{\sinh^2{v}}}\;\right]\,dv\right\}
\end{equation}
was found by Wilkins~\cite{Wil}. In fact, the asymptotic result
$N_n(-\infty,\infty)=\frac{2}{\pi}\log {n}$ was proven to be very
universal, i.e. insensitive to the details of the distribution of
coefficients $c_i$, see discussion and further references in
\cite{Far}. A similar universality result is also available for
the variance of the total number of real zeros~\cite{Mas}.

The nature of the informal arguments by Kac suggests that close to
the values $t=\pm 1$ many random terms of comparable magnitude add
up to form the polynomial, and the emerging object should be
universal in the spirit of the central limit theorem.

Another natural question arises about the expected number of real
zeros for coefficients with a nonvanishing mean value {\bf
E}$[c_i]=\mu$. It is straightforward to include the mean
value to the standard derivation of the Kac formula for gaussian-distributed
coefficients and arrive
at the following expression:
\begin{eqnarray}\label{Kac1}
&&
p_{n}^{\mu}(t)=p_{n}(t)
e^{-\frac{1}{2}\left[\Sigma_1(t)+\Sigma_2(t)\right]}
\int_0^{\infty} dq\;q \cosh{\left(q\sqrt{\Sigma_1(t)}\right)}
e^{-\frac{q^2}{2}}\\\
&& \Sigma_1(t)=\frac{A_n(t) D^2_n(t)}{A_n(t)C_n(t)-B_n^2(t)}\quad,\quad
\Sigma_2(t)=\frac{G^2_n(t)}{A_n(t)}
\end{eqnarray}
where
\begin{equation}
D_n(t)=-G_n(t)\frac{B_n(t)}{A_n(t)}+\frac{d}{dt}G_n(t)\quad,\quad
G_n(t)=\mu\sum_{j=0}^{n-1}t^{j}=\mu\frac{t^{n}-1}{t-1}
\end{equation}
and $P_n(t),A_n(t),B_n(t)$ and $C_n(t)$ are defined in
(\ref{Kac}).

Earlier investigation~\cite{Far} has revealed that any, whatever
small, mean  value $\mu>0$ asymptotically, for large $n$, kills
exactly half of real zeros and converts the above quoted result
\ref{Kacasy} for their number into
$N_n(-\infty,\infty)=\frac{1}{\pi}\log {n}$. It is this surprising
abruptness that motivated us to have a closer look at the origin
of such a behavior. This question turned out to be also related to
exploring the appropriately scaled vicinity of the accumulation
points.

\section{Scaling and Universality}

It is quite clear (and can be verified numerically) that the root
density function $p_n(t)$ as well as its limiting form
(\ref{limtriv}) are both not universal. On other hand, the essence
of the Kac argument discussed above and the mentioned universality
of $N_n(-\infty,\infty)$ suggests that universal features have to
emerge in close vicinity of the accumulation points $t=\pm 1$. In
fact, it is easy to understand that the relevant vicinity of
$|t|=1$ is of the order of $n^{-1}$ for large degree $n$. We will
refer to such domain as the \textit{'local scaling regime'}, as
opposed to keeping the distance $|1\pm t|$ from the accumulation points
$t=\pm 1$ fixed in the limit $n\to \infty$, the latter regime being
referred to as the \textit{'global'} one. In particular, our
main goal is to verify Proposition~\ref{asyprop}.
\begin{prop}\label{asyprop}
Let coefficients $\{c_j\}^{n-1}_{j=0}$ of a random polynomial
$f_n(t)$ be i.i.d real random variables with ${\bf E}
[c_j]=\mu_n>0$ and finite variance chosen to be unity: ${\bf
E}[c_j^2]=1$. Let  $N_n(x_1,x_2)$ be the number of real zeros of
$f_n(t)$ in the interval
$\left[1-\frac{x_2}{n}\,,\,1-\frac{x_1}{n}\right]$, with $ 0\le
x_1<x_2<n$. Then
\begin{eqnarray}\label{Th1}
\lim_{n\to\infty} {\bf
E}\left[N_n(x_1,x_2)\right]=\int_{x_1}^{x_2} p_{\alpha}(v) \,
dv
\end{eqnarray}
where $\alpha=\lim_{n\to \infty} (n\mu_n^2)$ and the density
function $p_{\alpha}(v)$ is given by
\begin{eqnarray}\label{scaleden}
&&
p_{\alpha}(v)=\frac{1}{\pi}\sqrt{\frac{1}{v^2}-\frac{1}{\sinh^2{v}}}
e^{-\alpha J(v)}\int_0^{\infty} dq\;q \cosh{\left(q\sqrt{\alpha
M(v)}\right)}
e^{-\frac{q^2}{2}}\\\
&& J(v)=\frac{\sinh^2{(v/2)}}{(v/2)^2}
\frac{1}{1+\sinh{v}/{v}}\quad\quad
M(v)=\frac{\sinh{v}/{v}}{\cosh^2{(v/2)}}
\frac{1-\sinh{v}/{v}}{1+\sinh{v}/{v}}
\end{eqnarray}
\end{prop}
A similar statement is valid for the vicinity of second
accumulation point $t=-1$ for $\mu_n<0$.

Here we present explicit arguments in favour of the validity of such a
proposition for the
simplest case $\mu_n=0$. The 'local regime' formula (\ref{scaleden})
reduces in that case to the following simple expression:
\begin{equation}\label{zeromean}
p_{0}(v)=\frac{1}{2\pi} \sqrt{\frac{1}{v^2}-\frac{1}{\sinh^2{v}}}
\end{equation}
The modifications required to include nonzero mean
$\mu_n$ are self-evident and left to the reader.

Our starting formula is the following representation for the
number of real roots $N_n(x_1,x_2)$ of $f_n(t)$ in the interval
$\left[1-\frac{x_2}{n}\,,\,1-\frac{x_1}{n}\right]$, see e.g.
\cite{Bleh,Far}:
\begin{equation}\label{1}
N_n(x_1,x_2)=\int_{1-\frac{x_2}{n}}^{1-\frac{x_1}{n}} \, dt
\int_{-\infty}^{\infty} dy\, |y|\, {\cal D}_n(0,y;t)
\end{equation}
where
\begin{equation}\label{2}
{\cal D}_n(x,y;t)=\mbox{\bf E}\left[\delta\left(x-\sum_{j=0}^{n-1}c_jt^j
\right)\delta\left(y-\sum_{j=0}^{n-1}jc_jt^{j-1}
\right)\right]
\end{equation}
stands for the joint probability density of the random polynomial
$f_n(t)$ and its derivative over $t$. Let us introduce in the
above representation the scaling variables $v$, $\tilde{x}$ and
$\tilde{y}$ by relations:
\[
t=1-v/n\qquad x=n^{1/2}\tilde{x}\qquad y=n^{3/2}\tilde{y}.
\]
Changing the variables of integration one finds:
\begin{equation} \label{3}
N_n(x_1,x_2)=\int_{x_1}^{x_2} \, dv \int_{-\infty}^{\infty}
d\tilde{y}\, |\tilde{y}| \, \tilde{{\cal D}}_n(0,\tilde{y};v)
\end{equation}
where
\begin{equation}\label{4}
\tilde{{\cal D}}_n(\tilde{x},\tilde{y};v)= \mbox{\bf
E}\left[\delta\left(\tilde{x}-\frac{1}{n^{1/2}}
\sum_{j=0}^{n-1}c_j(1-v/n)^j\right)
\delta\left(\tilde{y}-\frac{1}{n^{3/2}}\sum_{j=0}^{n-1}jc_j(1-v/n)^{j-1}\right)\right]\,.
\end{equation}

Evidently each of the scaled random variables $\tilde{x}$ and
$\tilde{y}$ is a sum of $n$ independent, although not identically
distributed terms, with the magnitude of fluctuations of each term
depending on the summation index $j$. Calculating the variances of
the scaled variables we find that they are given by
\begin{equation}\label{variance1}
A_n(v)=\mbox{\bf E}\left[\tilde{x}^2 \right]=
\frac{1}{n}\sum_{j=0}^{n-1}
(1-v/n)^{2j}=\frac{1-(1-v/n)^{2n}}{v(2-v/n)}
\end{equation}
and
\begin{equation}\label{variance2}
B_n(v)=\mbox{\bf E}\left[\tilde{y}^2 \right]=\frac{1}{n^2}
\sum_{j=0}^{n-1} j(1-v/n)^{2j-1}=\frac{1}{2n^2}\frac{d}{da}
\left(\frac{a^n-1}{a-1}\right)_{a=(1-v/n)^2}
\end{equation}
and their covariance is just
\begin{equation}\label{variance3}
C_n(v)=\mbox{\bf E}\left[\tilde{x}\tilde{y} \right]=
\frac{1}{n^3}\sum_{j=0}^{n-1}
j^2(1-v/n)^{2j-2}=\frac{1}{n^3}\frac{d}{da}\left[a\frac{d}{da}
\left(\frac{a^n-1}{a-1}\right)\right]_{a=(1-v/n)^2}\,.
\end{equation}
We further notice that all three quantities $A_n(v)$, $ B_n(v)$
and $ C_n(v)$ have a well-defined finite large $n$ limit:
\begin{equation}
A_n(v)\to A_{\infty}=\frac{1-e^{-2v}}{2v}\;\quad\qquad B_n(v)\to
B_{\infty}=-\frac{1}{2}\frac{dA_{\infty}}{dv}\;\quad\qquad
C_n(v)\to C_{\infty}=\frac{1}{4} \frac{d^2A_{\infty}}{dv^2}
\end{equation}
coinciding with the limiting values of $A,B,C$ in Eq.(\ref{Kac})
after rescaling $t=1-v/n$ and the limit $n\to \infty$ for a fixed
$v$. Now invoking the local central limit theorem~\cite{CLT} we
infer that the limiting joint probability density
$\tilde{D}_{\infty}$ of the scaled variables $\tilde{x}$ \&
$\tilde{y}$ tends when $n\to \infty$ to the normal law with
variances $A_{\infty},B_{\infty}$ and covariance $C_{\infty}$, as
if the polynomial coefficients were normal with unit variance.
Therefore, the Kac formula for the density of real roots should be
asymptotically valid in this case, and the formula
Eq.(\ref{zeromean}) immediately follows after substituting those
limiting values into the Kac expression Eq.(\ref{Kac}).
Separation of the two regimes and identification of the scaling is
in some sense similar to \textit{'local'} versus \textit{'global'}
scaling regimes in spectra of random matrices. Only when the
spectral parameters are scaled appropriately will the various
correlation functions characterising the eigenvalues of large
random matrices show a surprisingly robust universality
\cite{Deift,FSuniv}.

Although the expressions Eq.(\ref{scaleden}), and especially
Eq.(\ref{zeromean}) look very natural (cf. the structure of the
asymptotic result Eq.(\ref{Kacasy})) and their universality stems
from such basic fact as the central limit theorem,  we failed to
trace a similar statement in the available literature on random
polynomials with real coefficients. For complex zeros a kind of
'local' regime was studied in much detail by Shepp \&
Vanderbei~\cite{Shep} and Ibragimov \& Zeitouni~\cite{Ibr} who
showed that those zeros tend to concentrate asymptotically on the
unit circle. Only very recently (squared) expression of the type
(\ref{zeromean})
 independently emerged in studies of complex zeros of polynomials with
complex i.i.d. coefficients, see Shiffman and Zelditch,
\cite{Shif2}. A different kind of universality results was addressed in a
very recent paper \cite{Bleh1} which appeared when the present paper
was under completion.

\section{Numerical Examples, Discussion and Perspectives}\label{con}
\begin{figure}[h]
\begin{center}
\includegraphics[scale=0.4]{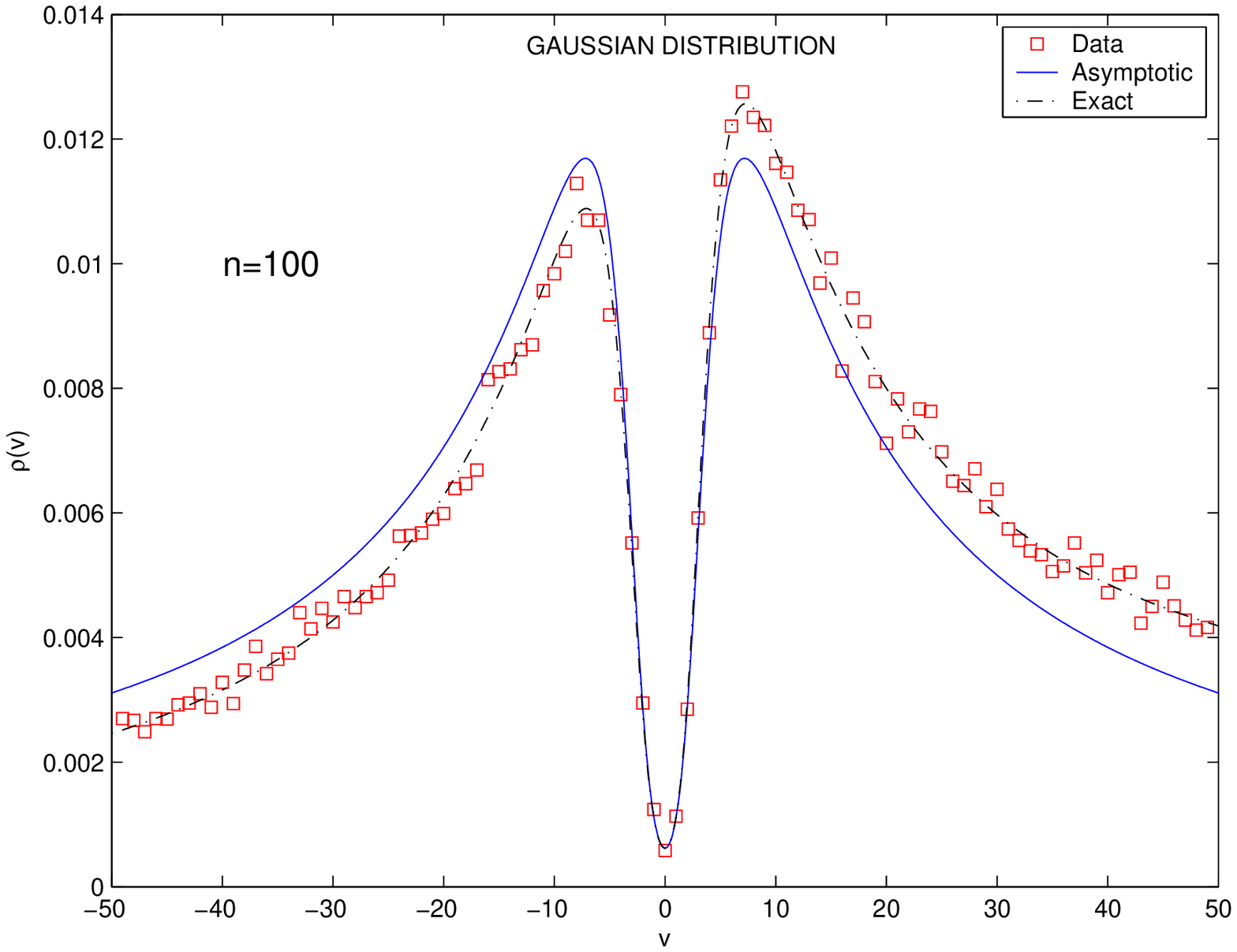}\;\;\;
\includegraphics[scale=0.4]{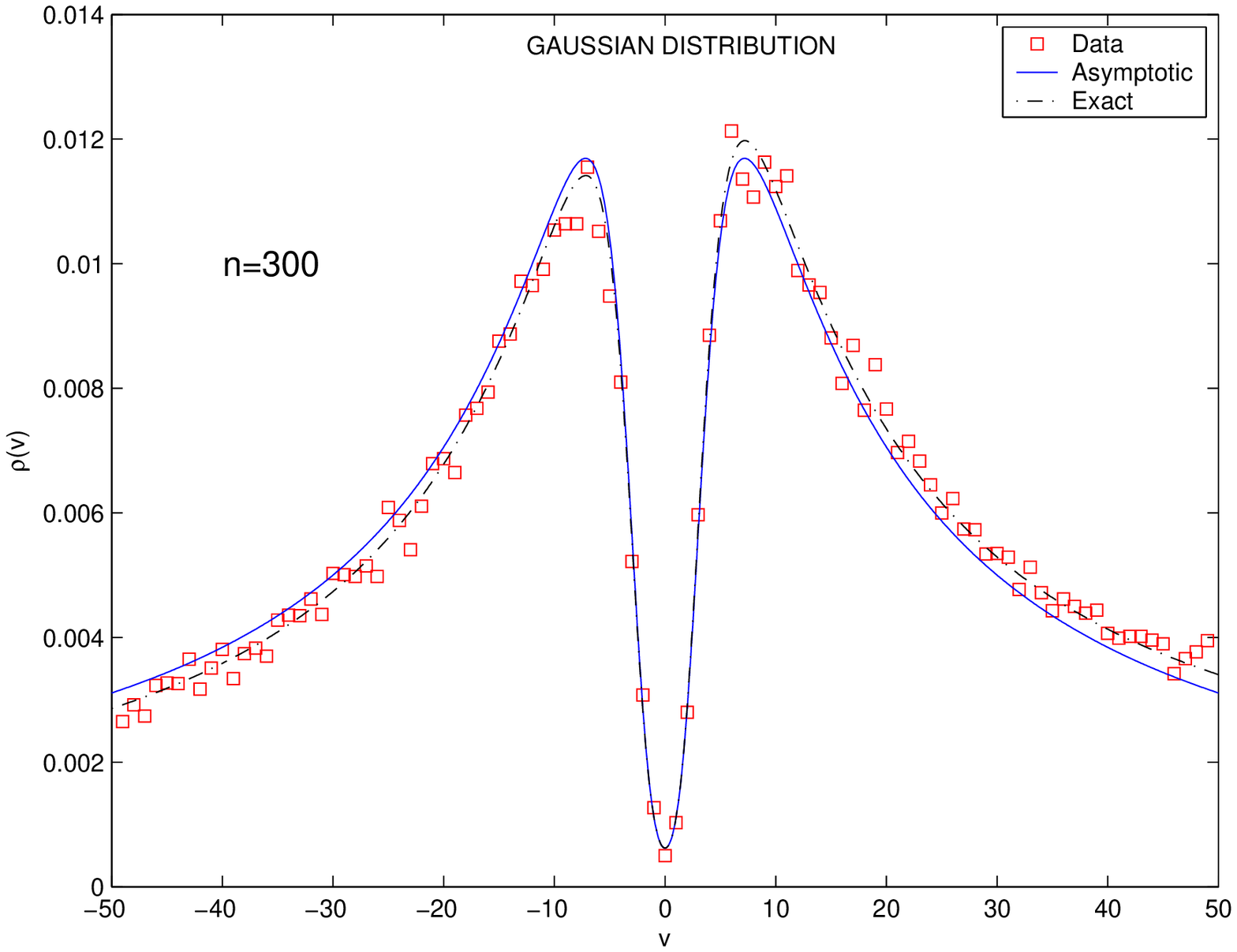}\;\;\;
\includegraphics[scale=0.4]{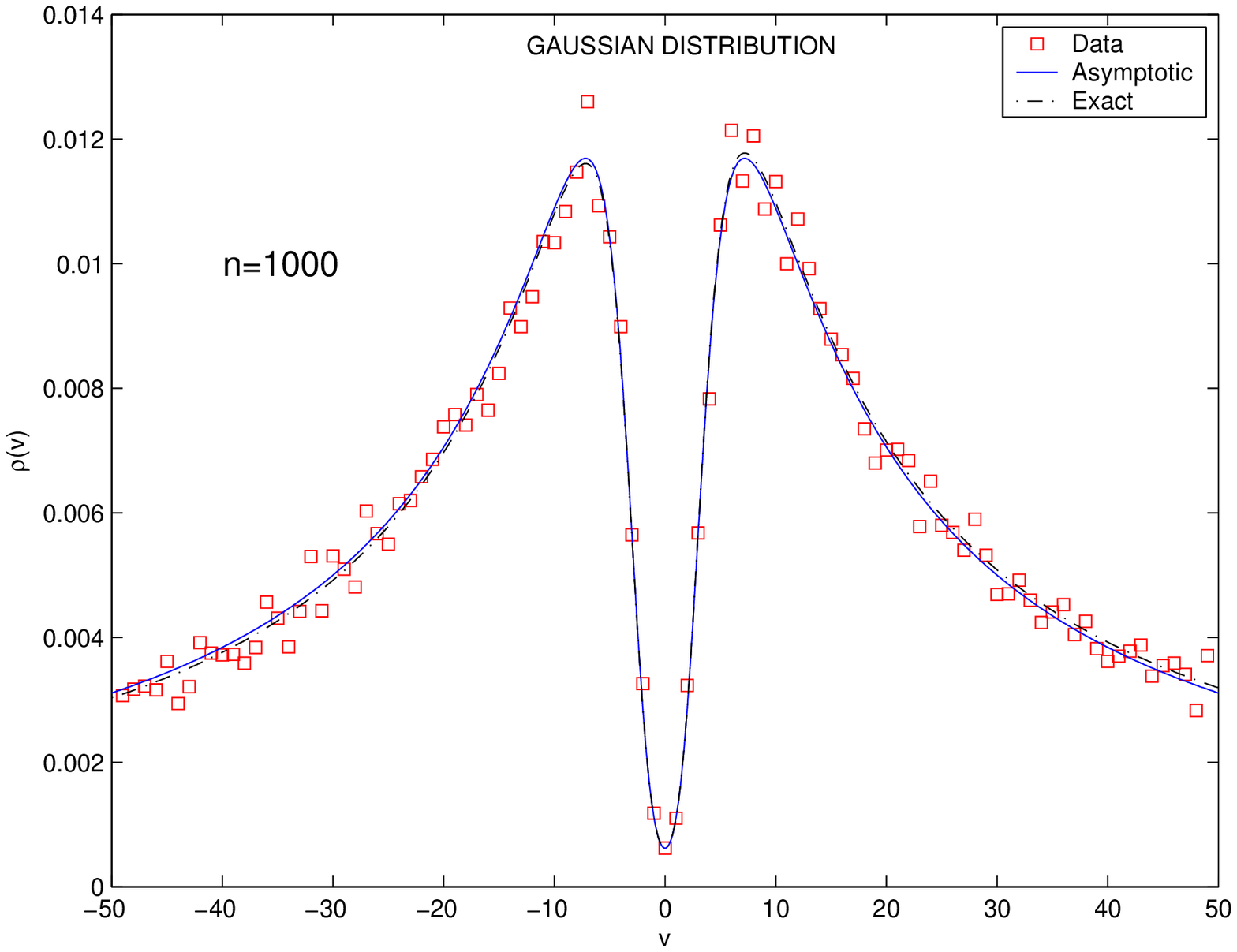}\;\;\;
\caption{Numerical vs. analytical density profiles for real roots
of polynomials with gaussian-distributed coefficients. Each data
set corresponds to $100,000$ realisations of random polynomials.
\label{GAUSSpics}}
\end{center}
\end{figure}

\begin{figure}[h]
\begin{center}
\includegraphics[scale=0.4]{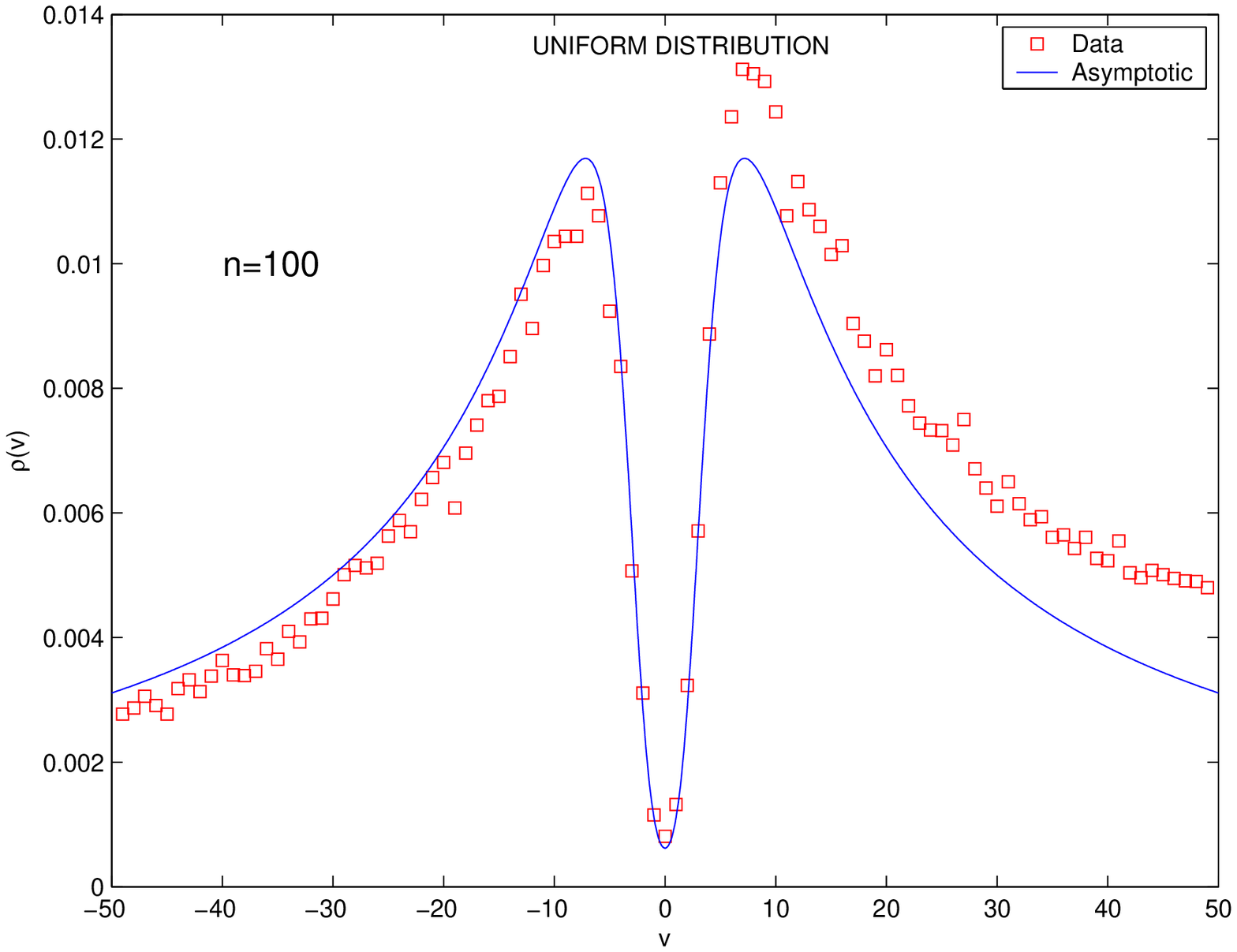}\;\;\;
\includegraphics[scale=0.4]{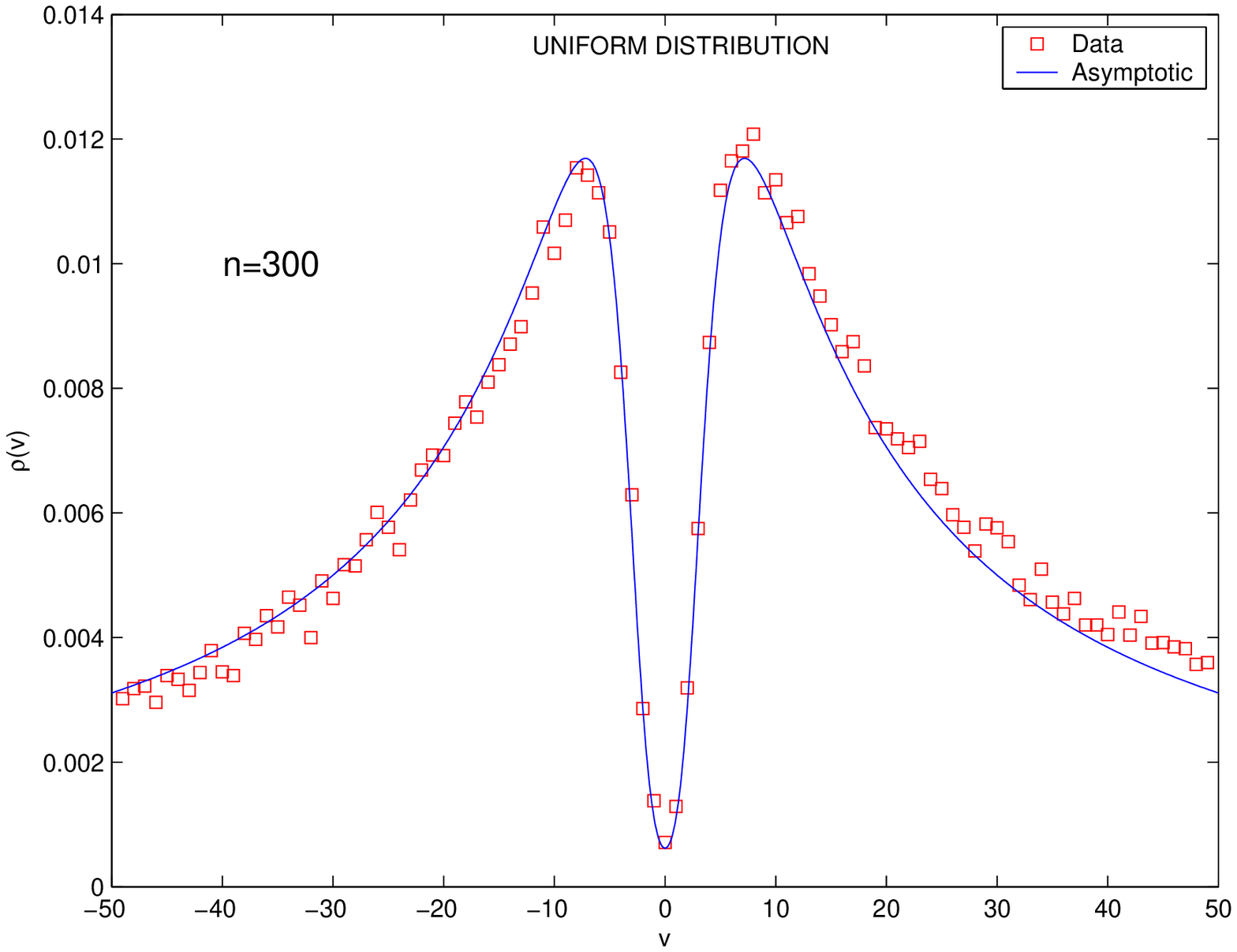}\;\;\;
\includegraphics[scale=0.4]{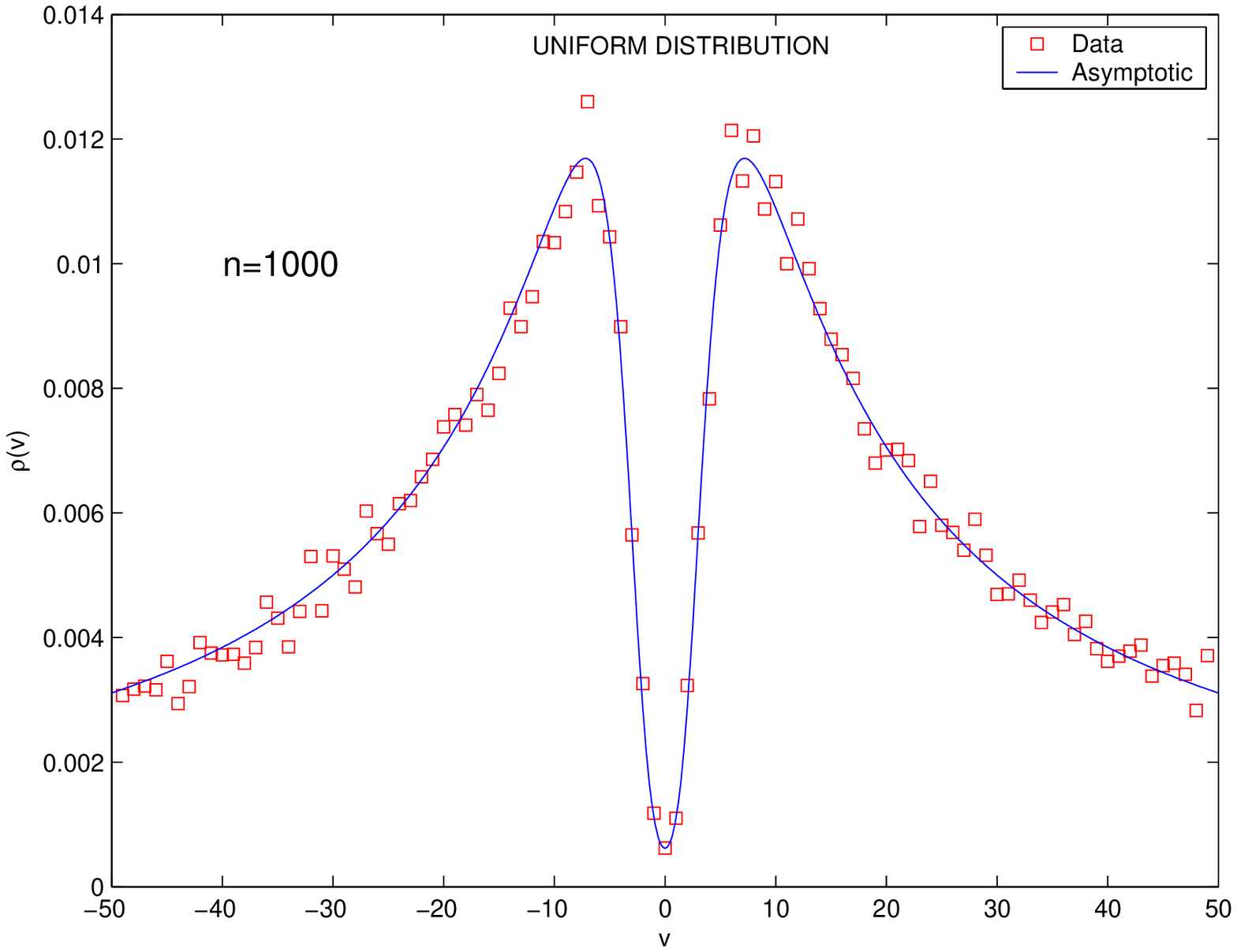}
\caption{Numerical data versus asymptotic density profile
(\ref{scaleden}) for the uniformly distributed coefficients. Each
data set corresponds to $100,000$ realisations of random
polynomials.} \label{UNIFORMpics}
\end{center}
\end{figure}

 It is interesting and informative to have a look at the profiles of the real root
density in the \textit{'local scaling regime'} around the point $t=1$
 (i.e $v=0$) for some typical values
of the parameters, and compare them with the results of numerical
simulations. In fig. (\ref{GAUSSpics})  we plotted the density
profiles obtained by a direct numerical search for real roots of
polynomials with gaussian-distributed coefficients for degrees
$n=100$, $300$ \& $1000$. The (scaled) mean value
$\alpha_n=\mu_n^2 n $ of the coefficients was chosen to be
$\alpha=10$ and the variance was always kept unity. The results
can be compared with both the exact predictions of the Kac-type
formula (\ref{Kac1}) for the same values of $n, \sigma$ and
$\mu_n=\sqrt{\alpha/n}$, and with the asymptotic universal profile
of (\ref{scaleden}).
It is trivial to see that as $n$ increases the exact Kac-like
expression begins to coincide with the asymptotic profile, and the
analytical curve agrees well with the numerical data. To verify
universality we also performed numerical simulations for the case
of coefficients uniformly distributed in the interval of the
widths $1/\sqrt{3}$ around the same mean values $\mu_n$. The
picture looks very similar, see fig. (\ref{UNIFORMpics}), and
again agreement with the asymptotic formula for large values of
$n$ is very good.

One can notice a couple of interesting features. First simple
observation is that the exact Kac formulae are asymmetric about
$v=0$ for 'small' $n$ (in the \textit{'local regime'}) but it
becomes "asymptotically symmetric" with increasing $n$, as
predicted by (\ref{scaleden}). This feature is easy to understand
in view of the global inversion symmetry $t\to t^{-1}$ which holds
exactly for all polynomials with i.i.d. coefficients. In the local
regime close to accumulation points this symmetry implies the
asymptotic reflection symmetry $v\to -v$.

The most surprising feature is a non-trivial double-peak structure
of the density profile, and it deserves to be discussed in more
detail.

\begin{figure}[h!!!]
\begin{center}
\includegraphics[scale=0.3]{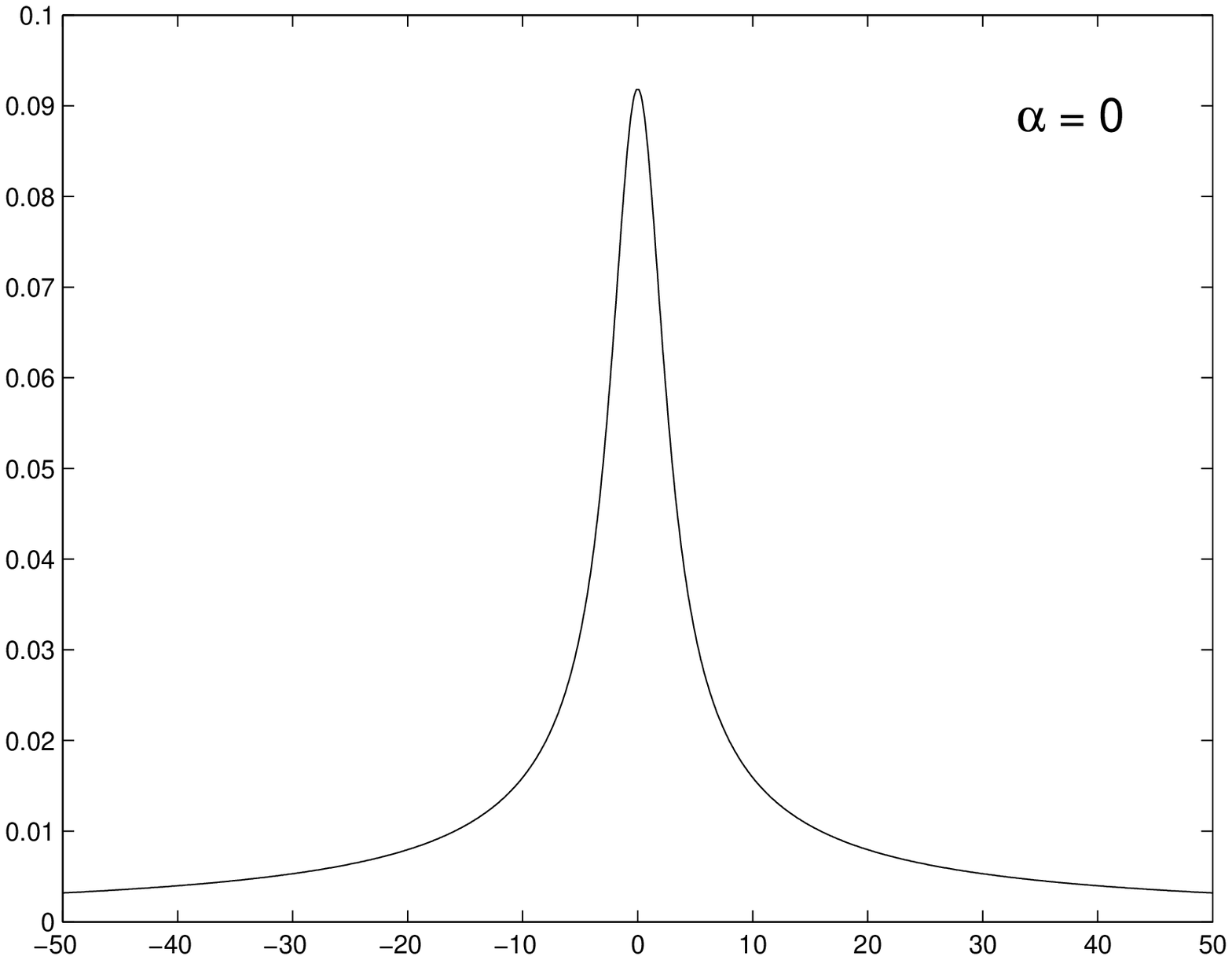}\;
\includegraphics[scale=0.3]{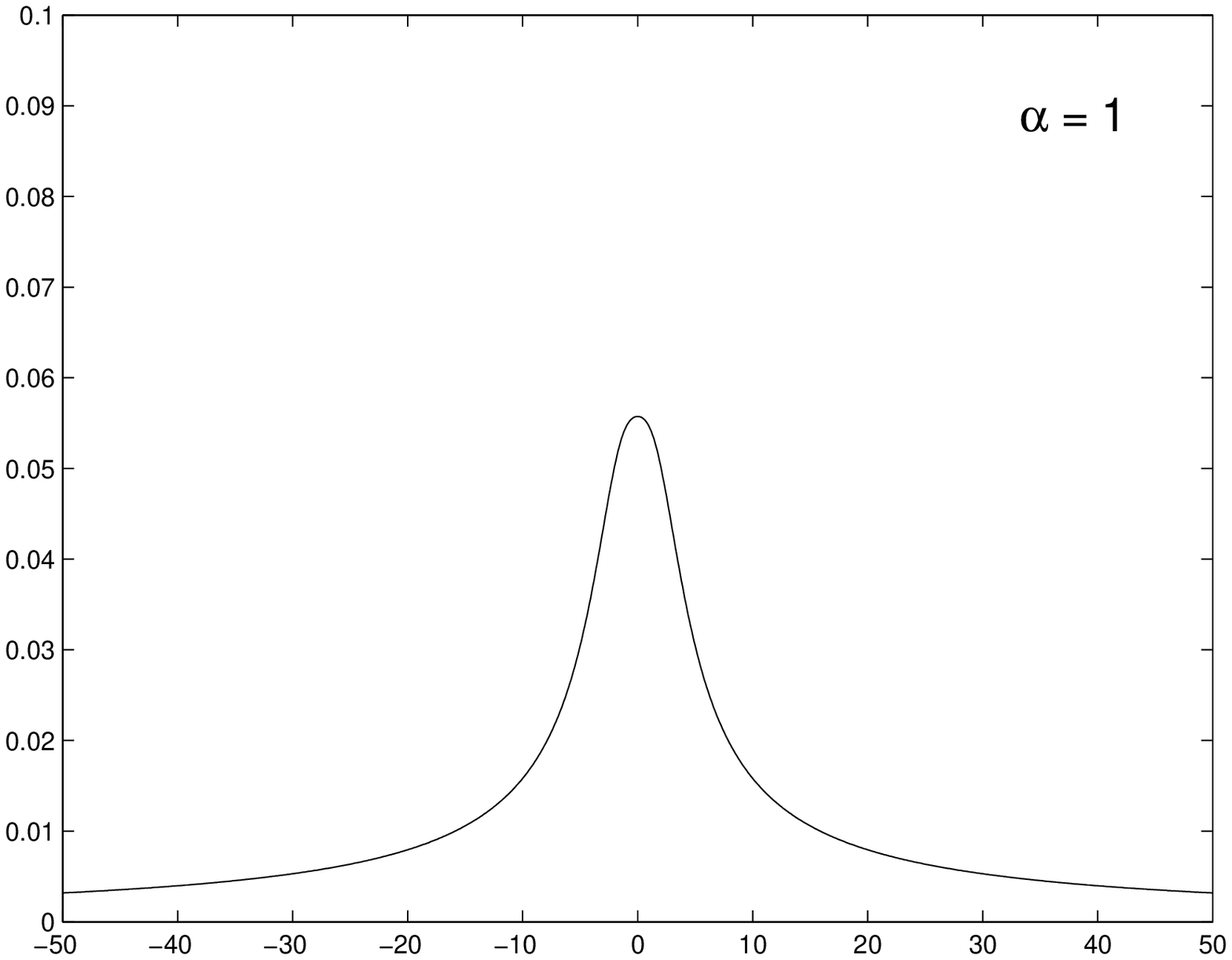}\;
\includegraphics[scale=0.3]{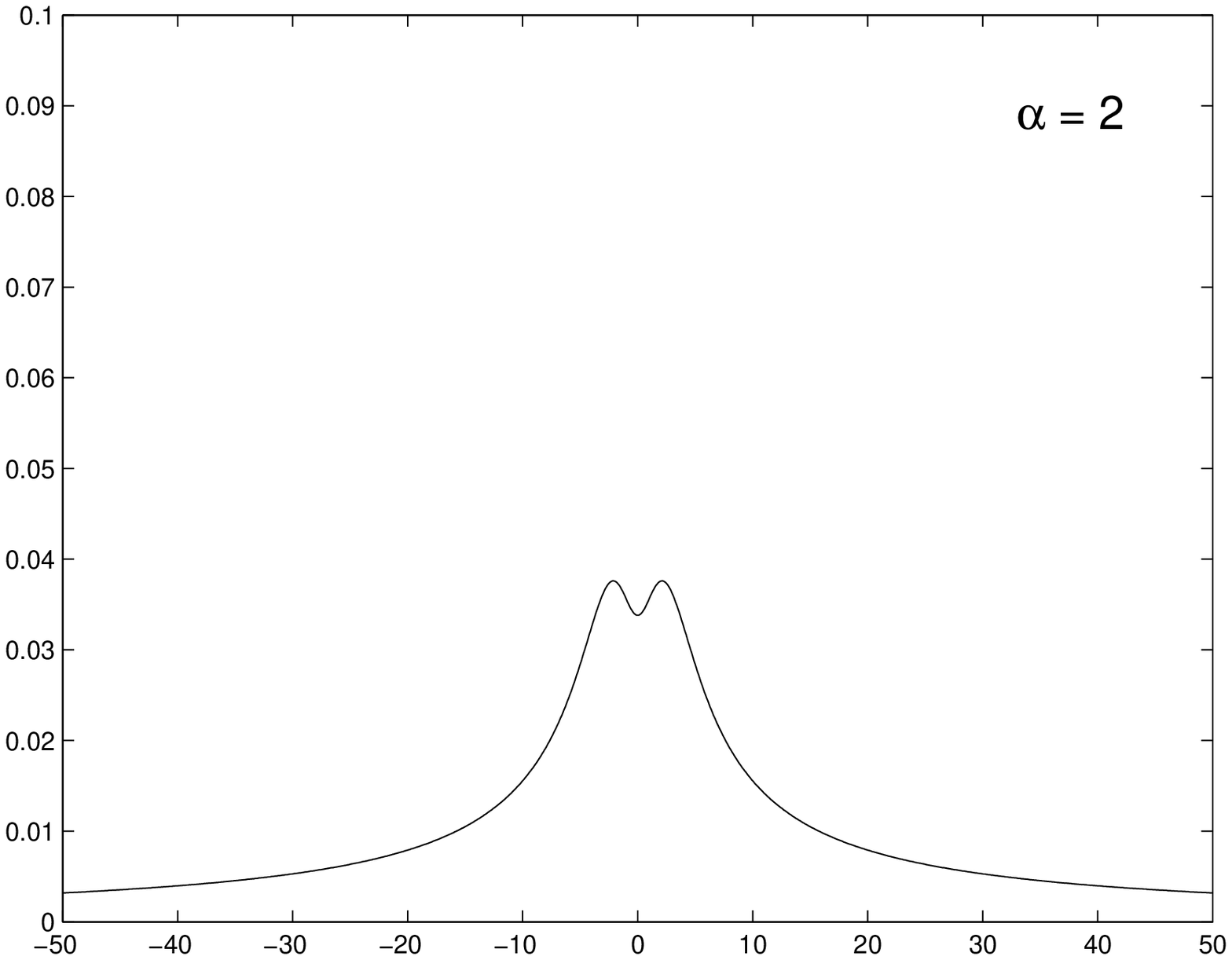}\;
\includegraphics[scale=0.3]{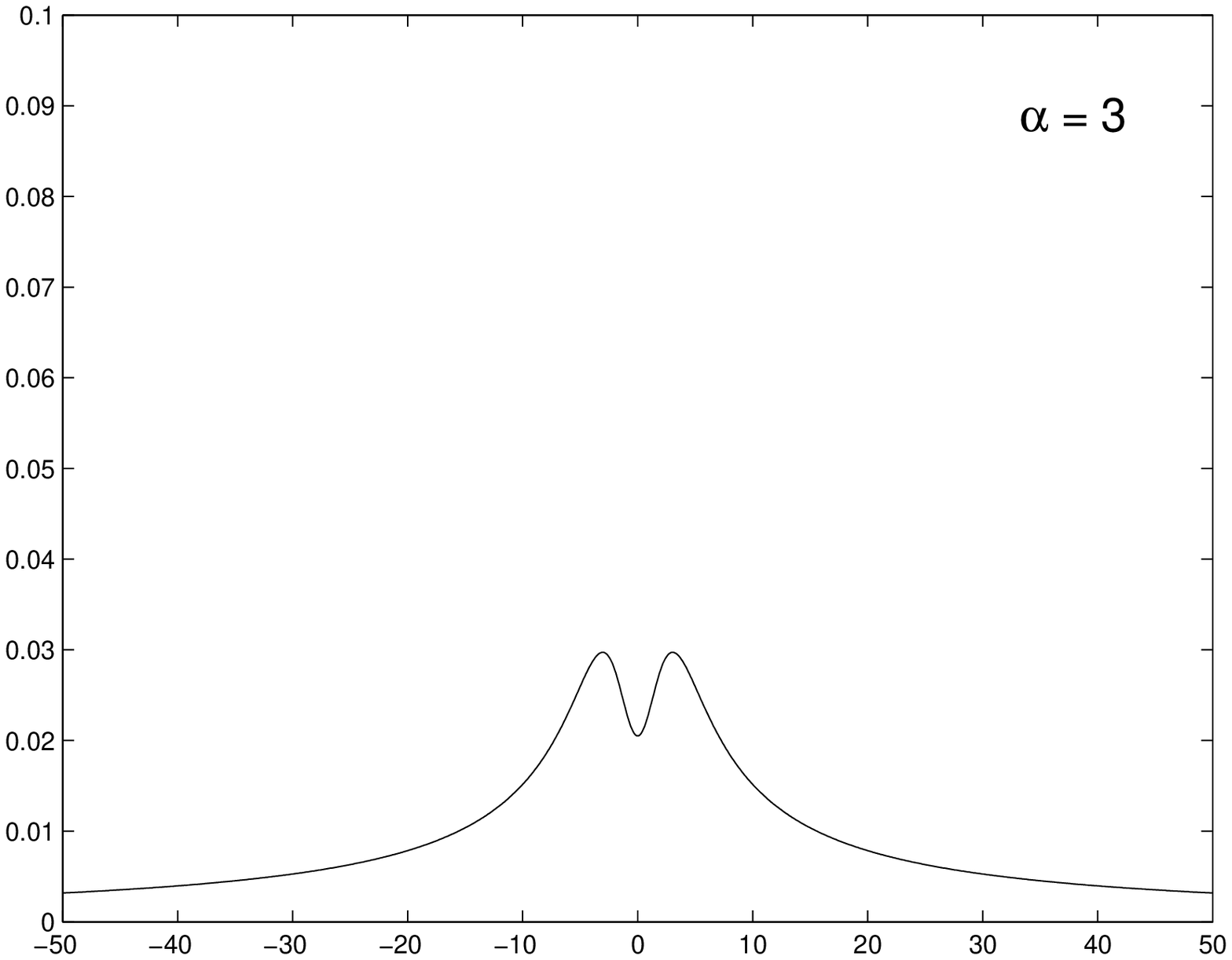}\;
\includegraphics[scale=0.3]{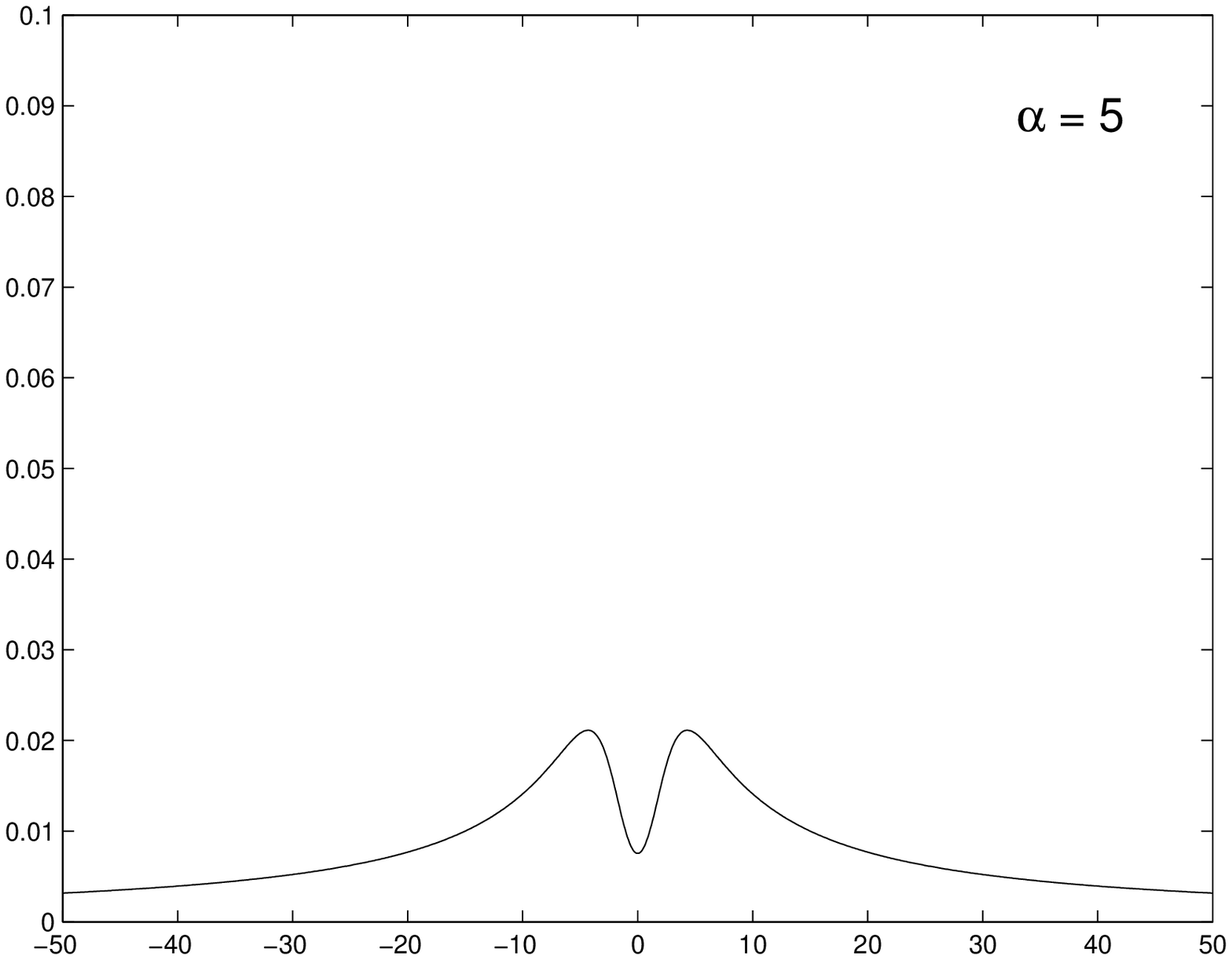}\;
\includegraphics[scale=0.3]{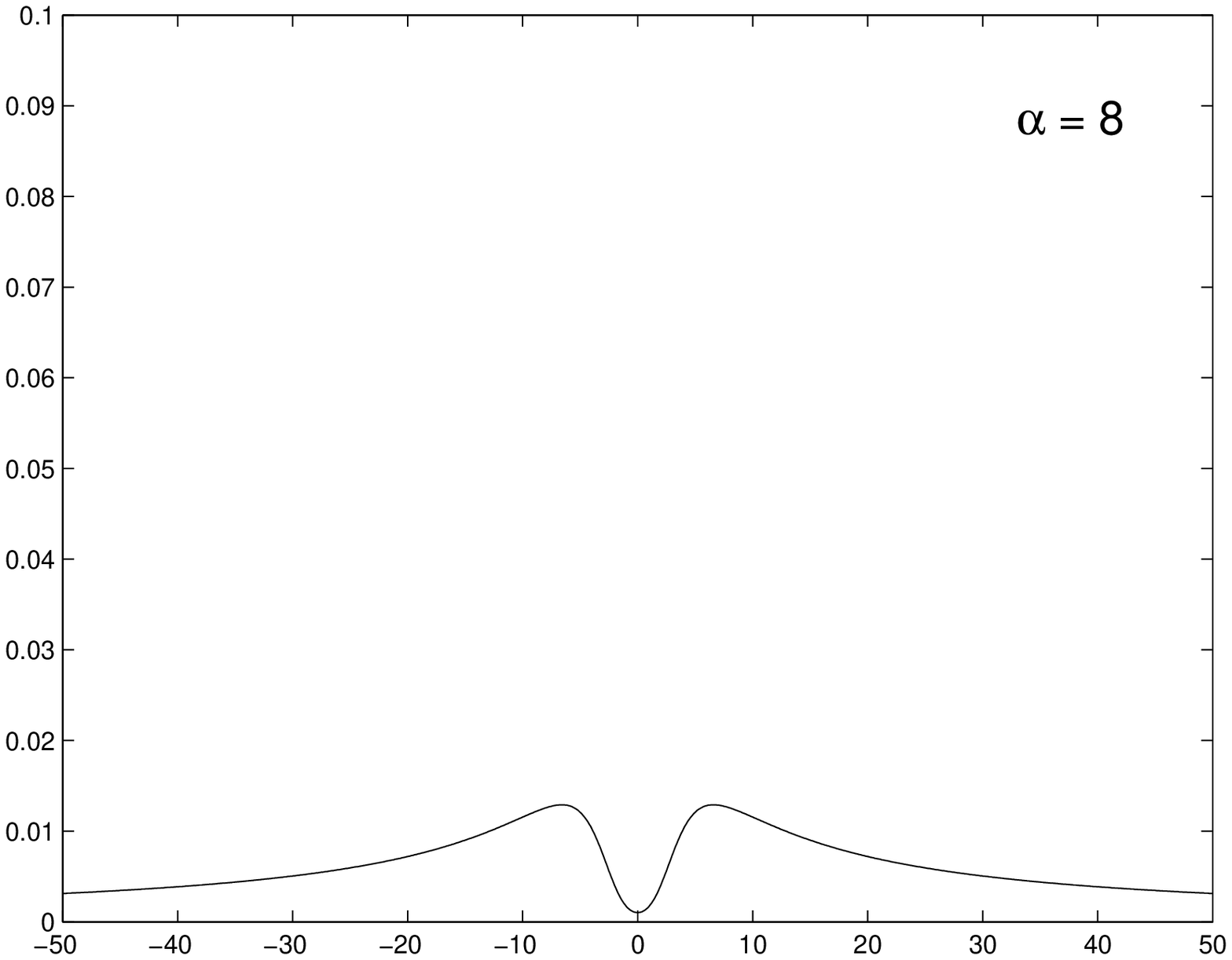}\;
\caption{Suppression of the density profile by increasing scaled
mean value $\alpha$ from $0$ to $8$ for a degree 100
polynomial.\label{alpha}}
\end{center}
\end{figure}

In fact, it turns out that as $\alpha$ increases the shape of the
density profile described by formula (\ref{scaleden})  changes
from that with one maximum to that with two symmetric maxima as
illustrated in the fig. (\ref{alpha}).

This fact can be verified analytically by considering the
small-$v$ expansion:
\begin{equation}\label{smallv}
p_{\alpha}(v)=\frac{1}{\pi \sqrt{3}} \;e^{-\alpha/2} \left[
1-\frac{v^2}{2}\left(\frac{1}{5}-\frac{1}{6}\;\alpha\right)
+\frac{v^4}{36}\left(\frac{137}{350}-\frac{5}{8}\;\al+\frac{1}{12}\;\al^2\right)
+O(v^5)\right]
\end{equation}
valid for $\alpha v^2\ll 1$. We see that for $\alpha>\alpha_c=6/5$
the local maximum at $v=0$ is converted to a minimum. On the other
hand, for $|v|\gg \alpha$ we easily find from Eq.(\ref{scaleden})
the $\alpha-$independent far tail decay law $p_{\alpha}(v)\approx
1/{\pi |v|}$ which ensures that at least two symmetric maxima have
to appear at $v\ne 0$. Notice that the far tail values are much
larger for $\alpha\gg 1$ than  the exponentially small density $
p_{\alpha}(0)=\frac{1}{\pi \sqrt{3}} \;e^{-\alpha/2} $
 at the origin, and thus the maxima have to be quite pronounced.

The precise reason for such a nonmonotonous behavior is not clear
to us at the moment, and deserve further investigations. We
consider that feature as an indication of a rather nontrivial
statistics of zeros accumulating in the scaling region around the
points $\pm 1$.

Fixing the value of $\mu$ and performing the limit $n\to \infty$
amounts to letting $\alpha\to \infty$. This results in vanishing
density of the real roots $p_{\alpha}(v)\to 0$ for any fixed $v$.
This fact corresponds to observed suppression of exactly half of
the real zeros (the vicinity of the second accumulation point is
not at all affected by the nonvanishing mean $\mu_n$). In
contrast, correctly tuning the mean value with the parameter $n$
by letting $\mu_n\sim n^{-1/2}$ and magnifying the vicinity of the
accumulation point via the correct scaling $1-t\sim n^{-1}$
results in a density function (\ref{scaleden}) and thus describes
a gradual (as opposed to abrupt) suppression of real zeros in the
scaling regime, see fig. (\ref{alpha}).

Let us now briefly discuss open questions and perspectives for
further research.

A very interesting development of the theory of random polynomials
comes from the paper by Bleher and Di~\cite{Bleh}. The authors
discovered nontrivial correlations between positions of real zeros
of algebraic polynomials. The correlations can be conveniently
characterised by correlation functions of the polynomial zeros:
\begin{equation}\label{bleh}
K(t_1,...,t_l)=
\lim_{n\to \infty}{\bf E}\left[
\lim_{\Delta_1,...,\Delta_l\to 0}\frac{N^{(f)}(t_1,t+\Delta_1)...
N^{(f)}(t_l,t_l+\Delta_l)}{|\Delta_1|...|\Delta_l|}\right]
\end{equation}
where $N^{(f)}(a,b)$ is the number of real zeros of the polynomial
 $f(t)$ in the interval $[a,b]$.

The consideration of correlation functions studied by Bleher and
Di was restricted, in our terminology, to the 'global regime'. It
should be possible to derive 'local regime' formulae for those
quantities. The expressions are expected to be universal in the
same sense as the 'local regime' formula for expected number of
roots, Eqs. (\ref{Th1}, \ref{scaleden}). The simplest nontrivial
quantity of that type should be the variance
$var{\left[N_n(x_1,x_2)\right]}$ of the number of real roots in
the vicinity of $t=1$.

Another natural object to study is the variation of the number of
real polynomial roots against a small change of the vector of real
coefficients ${\bf c}=c_0,...,c_{n-1}$. More specifically, change
${\bf c}\to {\bf c}_v={\bf c}+v{\bf b}$ assuming the components of
the vector ${\bf b}$ to be i.i.d. standard Gaussian. The parameter
$v\ge 0$ is used here to control the magnitude of the
perturbation. By using $f_v(t)$ to denote the perturbed polynomial
an interesting question arises, that is can we generalize the
correlation functions Eq.(\ref{bleh}) to the following {\it
parametric} correlation functions:
\begin{equation}\label{par}
K_{v_1,...,v_l}(t_1,...,t_l)=
\lim_{n\to \infty}{\bf E}\left[
\lim_{\Delta_1,...,\Delta_l\to 0}\frac{
N^{(f_{v_1})}(t_1,t+\Delta_1)...
N^{(f_{v_l})}(t_l,t_l+\Delta_l)}{|\Delta_1|...|\Delta_l|}\right]
\end{equation}
which reflects the change in the positions of the real roots.
Similar objects are of interest in the theory of random matrices
and disordered systems (see \cite{parametric} and references
therein).

As a preliminary step of our research we evaluated the simplest
nontrivial parametric correlation function, following the
paper~\cite{Bleh} and found in the 'local regime':
\[
\frac{K_{0,v}(t,t)}{[p(t)]^2}=
1+\frac{1}{v}\arcsin{\frac{1}{\sqrt{1+v^2}}}
\]
It will be interesting to attack the problem in full generality,
both for 'global' and especially for 'local' regime, where the
results are expected to be universal.

In fact, our initial interest in the properties of random
polynomials was stimulated by the fact that closely related
methods can be applied to study the properties of irregular
eigenfunctions in 'quantum billiards.' The eigenfunctions
$\Psi(x,y)$ at the point with the coordinate vector
$\vec{r}=(x,y)$ are solutions of the Helmholtz equation: $-\Delta
\Psi (\vec{r})=E \Psi (\vec{r})\quad; \quad \vec{r}\in \Omega$
where $\Omega$ is a connected compact domain with the boundary
$\partial \Omega$, and $\Delta$ is the Laplacian.

Recently Smilansky and collaborators~\cite{Smi} suggested looking
for the eigenfunctions at the point $\vec{r}$ in the following
representation:
\begin{equation}
\Psi(r,\theta)=\sum_{l=-L}^{L} a_l J_l(k|r|)e^{il\theta}\;
\end{equation}
where $J_l(x)$ stands for the Bessel functions, $r,\theta$ are
polar coordinates of the observation point, and the integer $L$ is
given in terms of the wavenumber $k$ and the perimeter $D$ of the
billiard boundary $\partial \Omega$ as $L=\frac{1}{2}\left[\frac{k
D}{\pi}\right]$. The complex coefficients $a_l$ satisfying
$\overline{a_l}=(-1)^l a_{-l}$ are taken to be i.i.d. complex
Gaussian variables with unit variance, in accordance to The
Berry's conjecture. Based on such a formula the authors of
\cite{Smi} managed to calculate the mean number $N$ and variance
$var(N)$ of the intersection of nodal lines with the billiard
boundary $\partial \Omega$. For the Dirichlet boundary conditions
along the boundary curve $\partial D$ parameterised as
$R(\theta)$, with $0\le \theta < 2\pi$, the problem turned out to
be equivalent to
 counting the number of real zeros of the function
\[
 u(\theta)=
\sum_{l=-L}^{L} a_l J'_l\left(kR(\theta)\right)e^{il\theta}
\]
in the interval $\theta\in [0,2\pi)$. Those developments provide
an interesting possibility for applying ideas and methods from the
theory of random polynomials to describe chaotic eigenfunctions.

Similar methods can be hopefully used to study sensitivity of the
nodal lines of eigenfunctions of 'quantum billiards' with respect
to the perturbation of the billiard parameters. For example, one
may wish to study parametric variations of the quantities
involved, with the role of the external parameter played by a
slight random variation of the boundary curve $R(\theta)$ or by
any other tunable physical parameter.

%
\section*{Acknowledgments}

This work was supported by EPSRC Doctoral Training Grant (APA) and
by Brunel University Vice-Chancellor Grant (YVF).


\indent
\end{document}